\documentclass[aps,prx,groupedaddress,twocolumn,showpacs,longbibliography,10pt]{revtex4-2}
\usepackage{hyperref}
\usepackage{amsmath}
\usepackage{amssymb}
\usepackage{graphicx}
\usepackage{slashed}
\usepackage{dcolumn}
\usepackage{bm}
\usepackage[inline]{enumitem}
\usepackage{xcolor}
\usepackage{dsfont}
\usepackage{bbold}
\usepackage{verbatim}
\usepackage{mathtools, xcolor,ytableau, amsfonts,tikz}
\usetikzlibrary{decorations.pathmorphing, decorations.pathreplacing,decorations.markings}
\usepackage{tabularx}

\newcommand{\tr}{\operatorname{tr}}

\usepackage{microtype}
\begin{document}

\preprint{APS/123-QED}

\title{Temperature-Resistant Order in 2+1 Dimensions}
\author{Zohar Komargodski}
\affiliation{Simons Center for Geometry and Physics, SUNY, Stony Brook, NY 11794, USA}

\author{Fedor K. Popov}
\affiliation{Simons Center for Geometry and Physics, SUNY, Stony Brook, NY 11794, USA}

\date{\today}

\begin{abstract}

High temperatures are typically thought to increase disorder. 
Here we examine this idea in Quantum Field Theory in 2+1 dimensions. For this sake we explore a novel class of tractable models, consisting of nearly-mean-field scalars interacting with critical scalars. We identify UV-complete, local, unitary models in this class and show that symmetry breaking $\mathbb{Z}_2 \to \emptyset$ occurs at any temperature in some regions of the phase diagram. This phenomenon, previously observed in models with fractional dimensions, or in the strict planar limits, or with non-local interactions, is now exhibited in a local, unitary 2+1 dimensional model with a finite number of fields.

\end{abstract}

\maketitle

\section{Introduction}

It is commonly assumed that fluctuations at high temperature disrupt the orderly patterns seen at lower temperature, leading to a state of greater entropy and disorder. This is justified by the free energy
\begin{equation}
F=E-TS
\end{equation}
which at high temperature is minimized by high-entropy states, which ought to be maximally disordered~\footnote{In the presence of imaginary chemical potentials, the entropy counts the states with phases and large cancellation can occur (unlike the ordinary thermal state). This leads to many examples where the high temperature states are ordered, see for instance~\cite{Komargodski:2017dmc,Dunne:2018hog} and references therein. A magnetic field may lead to similar issues~\cite{Farias:2021ult}.}.

One can make this line of reasoning precise under some assumptions. In lattice models with local interactions, the ultimate high temperature limit leads to a density matrix $\rho\sim \mathbb{1}$. This leads to disentanglement of the sites and forbids the formation of order, see e.g.~\cite{kliesch2014locality,Bakshi:2024cqr}. Some systems are known to exhibit a non-standard temporary ordered phase as they are heated -- e.g. transitions from a liquid to a solid as the temperature increases occur in He$_3$ (Pomeranchuk effect~\cite{lee1997extraordinary}), something similar occurs in bilayer graphene~\cite{rozen2021entropic}, and the Rochelle salt exhibits an enhanced order as it is heated~\cite{kao2004dielectric}.
The Rochelle salt also motivated S. Weinberg to construct a model in QFT with intermediate enhanced order~\cite{Weinberg:1974hy}. Here we will be interested in the true high temperature limit, and not in intermediate ordered phases.    

By contrast to lattice models, in Quantum Field Theory (QFT),  the notion of a unit density matrix does not exist (the unit operator is not a trace-class operator). Therefore it is not a priori obvious whether in QFT it is possible to have order persisting all the way to infinite temperature. 

Despite quite a bit of work, this problem in QFT remained open. In~\cite{Chai:2020onq,Chai:2020zgq} an example with order up to infinite temperature was constructed, but %it lived
in 3.99 dimensions and hence it was merely a toy model. (See therein many relevant references to older work on the subject.) Another such model was found in~\cite{Liendo:2022bmv}. Interesting attempts to clarify the situation using the AdS/CFT correspondence appeared in~\cite{Buchel:2020jfs,Buchel:2020thm,Buchel:2021ead,Buchel:2021yay,Buchel:2022zxl,Buchel:2023zpe}. Infinite temperature order was also recently discussed in non-local theories~\cite{Chai:2021tpt,Chai:2021djc}. Finally, interesting constructions with infinitely many fields were considered in~\cite{Chai:2020hnu,Chaudhuri:2020xxb,Bajc:2020gpa,Nakayama:2021fgy,Chaudhuri:2021dsq}. A closely related realization of a persistent confinement phase can be found in~\cite{Agrawal:2021alq}.  Spontaneous breaking of time-reversal symmetry has also been reported at high temperatures \cite{carlstrom2015spontaneous}. More recently, a general mechanism has been proposed to account for the persistence of spontaneous symmetry breaking across a variety of models \cite{Han:2025eiw}.

In this paper we consider a %slightly
new class of calculable models, consisting of scalars interacting with a large $N$ sector of critical scalars. In this model, one can rigorously establish the existence of symmetry breaking $\mathbb{Z}_2\to \emptyset$ at any temperature in 2+1 dimensions. Note that due to the Coleman-Hohenberg–Mermin–Wagner theorem, at finite temperature in a 2+1 dimensional theory we cannot have a broken continuous symmetry. Our result holds for finite, large $N$. Note the %recent 
paper~\cite{Hawashin:2024dpp}, which discussed the same model using the non-perturbative functional renormalization group, using a certain truncation method. The authors of~\cite{Hawashin:2024dpp} found persistent thermal order for $N>15$. Our contribution is to rigorously analyze the model for large, finite $N$ and demonstrate the phenomenon of persistent order. 

The field theory we construct is a multi-critical Conformal Field Theory (CFT) which breaks a $\mathbb{Z}_2$ symmetry at any temperature. That implies that in some regions of the phase diagram of this model symmetry breaking persists at all temperatures. It also follows that certain disordered (gapped) zero temperature phases become ordered as the temperature is increased. 
The idea of un-restored symmetry has had various applications in particle physics, see e.g.~\cite{Meade:2018saz,Bai:2021hfb,Babichev:2021uvl} and references therein for recent discussions of the possibility of un-restored electroweak symmetry. For the completeness of these discussions it would be nice to find examples of non-fine-tuned theories in 3+1 dimensions which exhibit persistent order. 

 It is not presently known if there are theories in 3+1 dimensions with finitely many degrees of freedom and local interactions that break a continuous symmetry at arbitrarily high temperature. Another open question concerns with the fact that our example is multi-critical, which means that persistent order is only a property of some high co-dimension slices of the phase diagram but may not be a generic property everywhere in the phase diagram. It would be nice to know if there are more generic theories with this property, and less fine tuned in the renormalization group sense. 
The existence of persistent order in QFT has clear implications for lattice models in the same universality class; we expect that in those lattice models the symmetries would be restored at temperatures of the order of the lattice scale, and much higher than the inverse correlation length, which is very unusual and would be nice to confirm. Moreover, symmetry breaking at high temperatures leads to deviations from the usual expectations for the density of heavy local operators -- while those deviations are mild in the case of a persistently broken $\mathbb{Z}_2$, they may be quite interesting for a persistently broken $U(1)$, see~\cite{Cao:2021euf,Barrat:2024aoa, Marchetto:2023xap, Benjamin:2023qsc} for recent literature on the density of heavy (bulk and defect) operators. Finally, our model could be extended to other setups where a weakly coupled CFT is coupled to a large $N$ CFT. Such models could provide a new class of theories with a tractable large $N$ expansion. For instance, one could study a scalar field interacting with a large-$N$ model coupled to a Chern-Simons theory at level $k$ \cite{Aharony:2011jz, Aharony:2012nh}, a free fermion coupled to a Gross-Neveu-Yukawa (GNY) model \cite{Moshe:2003xn} or even a free scalar field coupled to a large $N$ tensor model \cite{Klebanov:2016xxf}.

\section{The Setup}
The Wilson-Fisher family of fixed points is obtained in the infrared of $N$ real scalar fields with quartic interactions
$$S = \int d^{d} x \left[ \frac{1}{2} (\partial_\mu \phi_a)^2 + \lambda (\phi_a^2)^2 \right]~.$$
The coupling $\lambda$ is relevant in the ultraviolet for $d<4$ and the theory famously flows to a nontrivial CFT with $O(N)$ symmetry \footnote{We will consider only $3\leq d<4$}. 
This fixed point describes a zero temperature transition between an ordered phase with $N-1$ Nambu-Goldstone bosons and a trivial gapped phase.

The Hubbard–Stratonovich formalism allows studying the interacting fixed point more directly
\begin{gather}
    S = \int d^{d} x \left[ \frac{1}{2} (\partial_\mu \phi_a)^2 + \frac12 \sigma \phi_a^2  \right]~.
\end{gather}
We path integrate over $\sigma$ and over $\phi_a$.
At large, finite $N$, $\sigma$ is a primary operator of dimension slightly below $2$. Therefore, if we add a decoupled free scalar field $\psi$, of scaling dimension $d/2-1$, then the coupling $\sigma \psi^2$ would be a slightly relevant perturbation of the fixed point with a free field $\psi$. 
Therefore, let us consider the model
\begin{gather}\label{actone}
    S = \int d^{d} x \left[ \frac{1}{2} (\partial_\mu \phi_a)^2 + \frac12 \sigma \phi_a^2 + \frac12 (\partial_\mu \psi)^2 \right]+\delta S_{int}~,
\end{gather}
with
\begin{gather}\label{acttwo}
    \delta S_{\rm int} = \frac{t}{2}\int d^d x\, \sigma \psi^2~.
\end{gather}
For $3<d<4$ the model has no free parameters other than $t$, which flows to a fixed point. For $d=4$ one must however also consider operators such as $\sigma^2,\psi^4$, which are marginal and in $d=3$ one must also consider $\psi^6$ which is marginal. 
We will consider the $d=3$ case carefully. (Of course, for $d=4$, the model is infrared free.) In the range $3\leq d<4$  the fixed point would be multi-critical since $\sigma,\psi^2,\psi^4$ would all end up being relevant deformations.   

The way the large $N$ limit is utilized is by integrating out $\phi_a$ leading to
\begin{gather}
    S = \frac{N}{2}\tr \log\left[-\Delta + \sigma \right] +\int d^d x\left[\frac{1}{2} (\partial_\mu \psi)^2 + \frac{t}{2} \sigma \psi^2\right]
\end{gather}
We rescale $\sigma \to \frac{\sigma}{\sqrt{N}}$ to have an $O(1)$ two-point function in the large $N$ limit resulting in
\begin{gather}
    S = \frac{N}{2}\tr \log\left[-\Delta + \frac{\sigma}{\sqrt{N}} \right] +\int d^d x\left[\frac{1}{2} (\partial_\mu \psi)^2 + \frac{t}{2\sqrt{N}} \sigma \psi^2\right]~.
\end{gather}
Expanding the $\tr \log$ we find that the interaction vertex with $n$ insertions of $\sigma$ scales as $N^{2-n/2}$.
From that, we can immediately conclude that in the infinite $N$ limit $\sigma$ is a generalized free field of scaling dimension $2$. As we will see, corrections to the strict large $N$ limit are crucial to establish the existence of a unitary CFT. 

\section{The Large $N$ Limit}
The coupling $t$ is marginal at $N=\infty$ and slightly relevant at finite $N$. At large finite $N$ it therefore makes sense to discuss the beta function for $t$ since the flow is expected to be short. Since our utmost interest lies in $d=3$ we must also discuss the marginal operator $\psi^6$ and we denote its coupling by $g/N^2$ -- we will see that this is a natural scaling. 

The action that we would like to study is
\begin{align}
   & S = \frac{N}{2}\tr \log\left[-\Delta + \frac{\sigma}{\sqrt{N}} \right]  + 
   \cr &+\int d^d x\left[\frac{1}{2} (\partial_\mu \psi)^2 + \frac{t}{2\sqrt{N}} \sigma \psi^2 + \frac{g}{6! N^2} \psi^6\right]~.
\end{align}
Our strategy is to compute the beta function by considering contributions to the one-point functions $\langle \sigma \psi^2 \rangle$, $\langle \psi^6\rangle$.  
We start from $\beta_t$. We will see, self-consistently, that the contribution of the $g$ coupling to that beta function is highly sub-leading. 

The structure of $\beta_t$ is determined from conformal perturbation theory as (the ansatz below holds in $d=3$ only due to the absence of 
 a $t^2$ term, as explained shortly)
\begin{equation}\label{betaansatz}
\beta_t =\alpha {t\over N} +\beta {t^3\over N}+\cdots~.\end{equation}
The term proportional to $\alpha$ stems from the scaling dimension of the interaction $t$ in the $O(N)$ model coupled to a free field. That is why this term is of order $1/N$. The scaling dimension obeys $\Delta_\sigma =2 -{1\over N}\frac{4 \sin \left(\frac{\pi  d}{2}\right) \Gamma (d)}{\pi  \Gamma \left(\frac{d}{2}+1\right) \Gamma \left(\frac{d}{2}-1\right)}$, which in $d=3$ becomes $\Delta_\sigma = 2-{32\over 3\pi^2 }{1\over N}$. Therefore we ought to set $\alpha = -{32\over 3\pi^2 }$. Next, we need to explain why there is no term of order $t^2$ in~\eqref{betaansatz}. This is a miracle of $d=3$. Conformal perturbation theory shows that the coefficient of $t^2$ is controlled by the separated points three-point function $\langle \sigma\sigma\sigma\rangle/\sqrt {N}$. While in general dimension there is a cubic separated points $\sigma$ interaction of order $1/\sqrt N$, it is highly suppressed in $d=3$ and therefore can be dropped to leading order. (One can explain this miracle through a combination of parity and higher-spin symmetry. Note the analogy with the 2d Ising model, in which $\sigma$ flips sign under the Kramers-Wannier transformation, leading to a vanishing three-point function. Recently, it was further shown that the three-point function vanishes also at order $1/N^{3/2}$!~\cite{Goykhman:2019kcj}.)

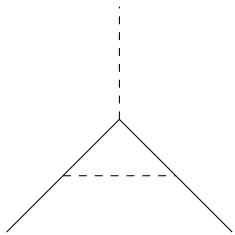
\begin{figure}
    \centering
    \begin{tikzpicture}[scale = 0.75]
        \draw[dashed] (0,0) -- (0,2);
        \draw (-2,-2) -- (0,0) -- (2,-2);
        \draw[dashed] (-1,-1) -- (1,-1);
    \end{tikzpicture}
    \caption{The diagram that gives the $t^3$ term in the beta function of $\sigma \psi^2$.}
    \label{fig:newt}
\end{figure}

Finally, the term proportional to $\beta$ in~\eqref{betaansatz} is related to the four-point function of the operator $\sigma\psi^2$, in which we use only Wick contractions to create a connected diagram and hence the $N$ scaling of the $t^3$ term, see figure~\ref{fig:newt}. 
Any higher order corrections would be suppressed in $1/N$, for instance, there is a contribution to the beta function of order $t^4/N^{3/2}$, but it can be dropped for now. 
Finally, we can determine $\beta$ without any new calculation. The idea is that 
$t=1$ has to be a zero of the beta function (at leading order in the large $N$ expansion)  since that corresponds to lumping $\psi$ together with the $\phi_a$ at a fixed point with $O(N+1)$ global symmetry. Therefore we find that   
\begin{equation}\label{cubicbeta}
\beta_t =-{32\over 3\pi^2 N}(t-t^3)+\mathcal{O}\left(N^{-\frac32}\right)%\cdots~.
\end{equation}

To verify that~\eqref{cubicbeta} is correct by a direct computation we need to make an important digression, which will be crucial also for the next step. 
While it is true that at separated points, to order $1/\sqrt N$, $\langle \sigma\sigma\sigma\rangle=0$, if one actually expands $\frac{N}{2}\tr \log\left[-\Delta + \frac{\sigma}{\sqrt{N}} \right]$ to cubic order in $\sigma$ one finds a term proportional to $\frac1{|p_1||p_2||p_1+p_2|}$ in momentum space. That leads to a contact term \cite{Arefeva:1978fj} %for 
$$\langle \sigma\sigma\sigma\rangle \sim {1\over \sqrt N}\delta^{3}(x-y)\delta^{3}(y-z)$$

%This contact term cannot be present at %finite $N$ since the scaling dimension %deviates from 2 at order $1/N$. %Therefore, the contact term represents %a separated points contribution in the %finite $N$ theory that just happens to %look like it is localized in position %space from the point of view of the %planar limit. 

This contact term is crucial to take into account in calculations, so let us determine it:
\begin{gather}
\vcenter{
\hbox{
\begin{tikzpicture}[scale=0.3]
\draw[dashed] (1,1) -- (-1,-1) node[near end, below left] {$p_1$}; % Left dashed line
\draw[dashed] (1,1) -- (1,3.5) node[near end, right] {$-p_1 - p_2$};     % Top dashed line
\draw[dashed] (1,1) -- (3,-1) node[near end, below right] {$p_2$}; % Right dashed line
 \draw[fill=black!20] (1,1) circle (0.5);
     % Right dashed line
\end{tikzpicture}}} = 
\vcenter{
\hbox{
\begin{tikzpicture}[scale=0.4]
% Left side: Dashed and wavy line
% Central triangle with wavy sides
\draw[decorate, decoration={snake}] (0,0) -- (1,1.73); % First side
\draw[decorate, decoration={snake}] (1,1.73) -- (2,0); % Second side
\draw[decorate, decoration={snake}] (2,0) -- (0,0); % Third side
% Dashed lines extending outward from the vertices
\draw[dashed] (0,0) -- (-1,-1) node[near end, below left] {$p_1$}; % Left dashed line
\draw[dashed] (1,1.73) -- (1,3) node[near end, right] {$-p_1 - p_2$};     % Top dashed line
\draw[dashed] (2,0) -- (3,-1) node[near end, below right] {$p_2$}; % Right dashed line
     % Right dashed line
\end{tikzpicture}}} = \\ = N \frac{(-1)^3}{N^\frac32} \frac{(-16 p_1) (-16 p_2) (-16 |p_1+p_2|)}{8 p_1 p_2 |p_1 + p_2|}
 =   \frac{512}{\sqrt{N}} \notag
\end{gather}
From that we conclude that we have the following contact term in the large $N$ limit
\begin{gather}
    \langle \sigma(x_1) \sigma(x_2) \sigma(x_3) \rangle = \frac{512}{\sqrt{N}} \delta^3(x_1-x_2) \delta^3(x_2-x_3)+\cdots
\end{gather}
where the $\cdots$ represent corrections of order $N^{-\frac32}$. 
Therefore, to verify the result~\eqref{cubicbeta} for $\beta_t$, the diagram that contributes to order $t^3$ is as in figure~\ref{fig:newt}, which we can compute easily
\begin{gather}
\vcenter{\hbox{
    \begin{tikzpicture}[scale = 0.4]
        \draw[dashed] (0,0) -- (0,2);
        \draw (-2,-2) -- (0,0) -- (2,-2);
        \draw[dashed] (-1,-1) -- (1,-1);
    \end{tikzpicture}
}}= 
    \frac{(-t)^3}{N} \int \frac{d^3 q}{(2\pi)^3} \frac{(-16 q)}{q^4} =  \frac{8 t^3}{\pi^2 N^\frac32} \log \Lambda
\end{gather}
while the diagram that has the cubic $\sigma$ vertex $\sim {1\over \sqrt N }\frac{\sigma(p_1)\sigma(p_2)\sigma(-p_1-p_2)}{|p_1||p_2||p_1+p_2|}$ (denoted by the gray filled circle below) 
\begin{gather}
    \begin{tikzpicture}[scale = 0.4]
        \draw (-2,0) -- (2,0);
        \draw[dashed] (-1,0) -- (0,2);
        \draw[dashed] (1,0) -- (0,2);
        \draw[dashed] (0,2) -- (0,3.5);
        \draw[fill=black!20] (0,2) circle (0.25);
    \end{tikzpicture}
\end{gather}
does not give logarithmic contributions at order $1/N$.
Finally, we need to take into account corrections to the $\psi$ propagator
\begin{gather}
\vcenter{\hbox{
    \begin{tikzpicture}[scale = 0.5]
        \draw (-2,0) -- (2,0);
        \draw[dashed] (-1,0) .. controls (0,1) .. (1,0);
    \end{tikzpicture}
    }} 
    % + \vcenter{\hbox{
    % \begin{tikzpicture}[scale = 0.5]
    %     \draw  (-2,0) --node[midway, above] {$-p^2 \delta_Z$}  (2,0)  ;
    %     \draw[fill=black!20] (0,0) circle (0.25);
    % \end{tikzpicture}
    % }} 
    = \notag\\
% \end{gather}
% This integral is equal to
% \begin{gather}
     \frac{(-t)^2}{N} \int \frac{d^3 q}{(2\pi)^3} \frac{(-16) q}{(p+q)^2} = 
     %- \frac{4 t^2}{N} \int \frac{q^3 dq dx }{\pi^2} \frac{1}{p^2 + 2 p q x + q^2} = \notag\\
    % - \frac{4 p^2 t^2}{N} \int \frac{dq}{\pi^2 q} \int\limits^1_{-1} dx (4 x^2 - 1) =
    - \frac{8 p^2 t^2}{3\pi^2 N} \log \frac{\Lambda}{p} 
\end{gather}
% And we choose counter term to be
% \begin{gather}
%     \delta_Z =  \frac{8 t^2}{3\pi^2 N}\log \frac{M}{\Lambda},
% \end{gather}
Hence, we find the anomalous dimension of the field $\psi$
% Applying the Callan-Symanzik equation we obtain the  anomalous dimensions of the field $\psi$
% \begin{gather}
%     \left(M \frac{\partial}{\partial M} + 2\gamma_\psi\right) \left(\frac{1}{p^2} - \frac{8 t^2}{3\pi^2 p^2 N} \log \frac{M}{p} \right)  = 0 \notag\\
%     \gamma_\psi = \frac{4 t^2}{3\pi^2 N }, \quad 
\begin{gather}
\Delta_\psi = \frac12 + \frac{4 t^2}{3\pi^2 N}~.
\end{gather}
At nonzero $t$ this can effectively change the renormalization group evolution and therefore we need to add $\frac{8 t^3}{3\pi^2}$ to the beta function of $t$. In summary we find 
\begin{gather}
    \beta_t =  \frac{8 t^3}{3\pi^2 N} + \frac{8 t^3}{\pi^2 N } - \frac{32}{3\pi^2 N} t
\end{gather}
exactly as anticipated in~\eqref{cubicbeta}. At general $d$ one can find that
 \begin{gather}\label{gendfx}
    \beta_t = -\frac{2^{d-1} (t-1) t \left(d^2-3 d+2 t+2\right) \sin \left(\frac{\pi  d}{2}\right)
   \Gamma \left(\frac{d}{2}-\frac{1}{2}\right)}{\pi ^{3/2} \Gamma
   \left(\frac{d}{2}+1\right) N}
\end{gather}
(note that now there is a $t^2$ term) 
The beta function~\eqref{gendfx} has zeros at $t=0,1$ (describing the $O(N)$ model with a decoupled free scalar and the $O(N+1)$ model, respectively) and $t = -\frac{(d-1)(d-2)}{2}$, which is an interesting nontrivial fixed point. 

At $d=3$ our main interest will be at the fixed point with $t=-1$ of~\eqref{cubicbeta}. To analyze it we have to complete the analysis of the sextic interaction $\psi^6$ and compute $\beta_g$. 
There are contributions from a dozen diagrams which we evaluate in Appendix~\ref{A}. After the dust settles we find \footnote{We thank Soumyadeep Chaudhuri for  verifying this beta function.}
\begin{gather}\label{gcoupling}
    \beta(g) = \frac{128 g t^2}{\pi^2 N} - \frac{245760}{\pi^2 N } t^3(t-1)^3,
\end{gather}
We can now discuss the fixed points. From~\eqref{cubicbeta}
we find the fixed points 
$t=\pm 1,0$. $t=0$ is a theory of a decoupled free scalar field along with the $O(N)$ model. For $t=1$ we find that also $g=0$ and we thus have the usual $O(N+1)$ vector model. Lastly, we find the solution of most interest to us here
\begin{gather}\label{gcouplingi}
    t = - 1, \quad  g = 15360~.
\end{gather}
It is crucial that $g>0$ and hence this is a unitary fixed point, with a bounded from below potential and no flat directions. This fixed point exists at large finite $N$, as our calculations demonstrated. We next study the effective potential at zero and nonzero temperature.

\section{Thermal Partition Function}
 We will discuss the effective potential starting from the model in $3<d<4$, where the fixed points are the solutions of~\eqref{gendfx}. 
After this warm-up we will return to the model of interest in $d=3$ with the sextic $\psi$ interaction. 

The action is as in~\eqref{actone},~\eqref{acttwo}, which we present again for convenience
\begin{gather}
    S = \int d^d x \left[\frac12 \left(\partial_\mu \phi_i\right)^2 + \frac12 \sigma \phi_i^2  + \frac12 \left(\partial_\mu \psi\right)^2 +  \frac{t}{2} \sigma \psi^2 \right], \label{eq:themodel}
\end{gather}
This theory has many flat directions classically. For instance, constant $\sigma$ and vanishing $\psi,\phi_i$ costs no energy classically. If $t$ is negative (as for the fixed point of interest here) we can also take $\phi_i^2=-t\psi^2$
with arbitrary $\sigma$.
We will compute the quantum corrections and the thermal corrections to find the correct minimum of the free energy.  

If $\phi_i$ has a constant VEV, we can, with no loss of generality, assume that only $\phi_1\equiv \phi$ is nonzero. Thus, after integrating the other components, we arrive at 
\begin{gather}
    S_{\rm eff, 0}[\sigma, \psi] = \frac{N - 1}{2} \tr \log \left(-\Delta +\sigma\right) + \notag\\
    +\frac12 (\partial_\mu \psi)^2 + \frac12 (\partial_\mu \phi)^2 + \frac12 \sigma\left(\phi^2 + t \psi^2\right) ,
\end{gather}
Let us consider the vacuum state in the large $N$ limit. We rescale $\phi \to N^\frac12 \phi, \psi \to N^\frac12 \psi$ for convenience. Using
\begin{gather}
    \frac{1}{2} \int \frac{d^d k}{(2\pi)^d} \log\left(k^2 +\sigma\right) =C_0 +  r_0 \sigma -\frac{\Gamma(-d/2)}{2(4\pi)^\frac{d}{2}} \sigma^{d/2}
\end{gather}
The effective potential is (we have thrown away the linear term in $\sigma$ by fine tuning to the fixed point)  
\begin{gather}
   \frac{1}{N} S_{\rm eff,0}[\sigma, \phi] = -\frac{\Gamma(-d/2)}{2(4\pi)^\frac{d}{2}} \sigma^{d/2} + \frac12 \sigma(\phi^2 + t \psi^2)
\end{gather}
We need to find the extrema of this action as a function of $\phi,\psi$ and $\sigma$. Varying with respect to $\sigma$ and $\phi,\psi$ we get the following equations
\begin{gather}
\frac{\Gamma\left(1-d/2\right)}{2(4\pi)^\frac{d}{2}} \sigma^{\frac{d}{2} - 1} + \frac12 \phi^2 + \frac{t}{2}\psi^2 = 0, \quad 
\sigma \phi = \sigma \psi = 0, \notag\\
 \phi^2 + t \psi^2 = 0~.
\end{gather}
For $t=1$, i.e. the $O(N+1)$ invariant fixed point, the only solution is $\sigma=\phi=\psi=0$. This is just the familiar statement that the $O(N+1)$ Wilson-Fisher conformal field theory has a unique vacuum. For $t=0$ we have $\phi=\sigma=0$ and $\psi$ is arbitrary. The arbitrary value of $\psi$ just reflects the vacuum degeneracy of the decoupled free field $\psi$. 

For $t<0$ we have a continuum of solutions,
$\sigma = 0$ and $\phi^2 = -t \psi^2$. In particular, we see a continuum of solutions at the fixed point $t = -\frac{(d-1)(d-2)}{2}$, which solves~\eqref{gendfx}. Flat directions are common in supersymmetric theories, here the flat direction is only a property of the theory at the leading nontrivial order in the  $\frac 1 N$ expansion. Interesting recent discussions about flat directions in non-SUSY theories can be found in~\cite{Nogradi:2021zqw,Cuomo:2024vfk,Cuomo:2024fuy}. 

To ascertain that we are dealing with a healthy conformal theory with an isolated vacuum, we should take into account the  $\frac{1}{N}$ corrections to the effective potential and find the true minimum of the free energy of this system. For that we expand our action to the second order in fluctuations and derive the effective potential from integrating out the fluctuations at one loop:
\begin{gather}
    % \phi = \phi + \delta\phi, \quad \psi = \psi + \delta\psi, \quad \sigma = \frac{1}{\sqrt{N}}\delta \sigma \notag\\
    % S = \int \frac{d^dp}{(2\pi)^d}\left[ \frac{p^2}{2} \delta\phi(p) \delta\phi(-p) + \frac{p^2}{2} \delta\psi(p) \delta\psi(-p) + \right.\notag\\ \left.
    % + \frac{ G^{-1}_\sigma(p)}{2} \delta\sigma(p)  \delta\sigma(-p) +  \phi \delta\sigma(p) \delta\phi(-p) +  t \psi  \delta\sigma(p) \delta\psi(-p)  \right], \notag\\
    S_{\rm eff,1}[\phi,\psi] = \frac12 \int \frac{d^d p}{(2\pi)^d} \log\left[- p^4 G^{-1}_\sigma(p)+ p^2 \left(\phi^2 + \psi^2 t^2\right)  \right]  \notag\\
    = \frac{\pi ^{d/2} \Gamma \left(-\frac{2}{d-2}\right) \Gamma \left(\frac{d}{d-2}\right)
   }{\Gamma \left(\frac{d}{2}+1\right)} \times \notag\\
   \times \left(-4^{d-1} \pi ^{\frac{d-3}{2}} \sin \left(\frac{\pi  d}{2}\right) \Gamma
   \left(\frac{d-1}{2}\right)\right)^{\frac{d}{d-2}} (\phi^2 + \psi^2 t^2)^\frac{d}{d-2}
    \notag~. 
\end{gather}
$G^{-1}_\sigma(p)$ comes from expanding $N\tr \log\left[-\Delta + \frac{\delta\sigma}{\sqrt{N}} \right]$ to second order in $\sigma(p)$ \cite{Moshe:2003xn}. The coefficient in front of  $\phi^2 + t^2  \psi^2$ is positive and therefore the origin is the only vacuum at zero temperature at the critical point .  
We see that at $d=3$ there is a logarithmic divergence in the effective potential which is, of course, related to the existence of a marginal operator. 

%Before we return to $d=3$ 
Let us study the situation at finite temperature of the model \eqref{eq:themodel} in $3<d<4$. The leading order effective potential is
\begin{gather}
    \frac{1}{N} S_{\rm, \beta} = \frac{1}{2\beta} \sum_n \int \frac{d^{d-1}k}{(2\pi)^{d-1}} \log\left(\omega_n^2 + k^2 + \sigma\right) + \notag\\
    + \frac12 \sigma \left(\phi^2 + t \psi^2\right)
\end{gather}
We again get that 
\begin{gather}
    \frac{\delta S}{\delta \phi} = \sigma \phi = 0, \quad   \frac{\delta S}{\delta \psi} = \sigma \psi = 0, \label{eq:largeNeq34d}
\end{gather}
and for the $\sigma$ equation, we find
\begin{gather}
\frac{1}{2\beta} \sum_n \int \frac{d^{d-1}k}{(2\pi)^{d-1}} \frac{1}{\omega_n^2 + k^2 + \sigma} 
    + \frac12 \left(\phi^2 + t \psi^2\right) = 0 \notag
\end{gather}
we pick $\sigma = 0$, which translates to
\begin{gather}
- t \psi^2 - \phi^2 = \frac{\Gamma\left(\frac{d-2}{2}\right)\zeta(d-2)}{4 \pi^\frac{d}{2} \beta^{d-2}} > 0~.
\end{gather}
As before, for positive $t$ there is no flat direction and there is a unique minimum of the free energy at the origin, both at zero and nonzero temperature. 
For negative $t$, 
the zero temperature flat direction is deformed. However, the flat direction is not lifted! Therefore, to leading order in $\frac1N$, the theory has a flat direction at zero temperature and a deformed flat direction at nonzero temperature. Taking into account the $\frac{1}{N}$ corrections we obtain a small correction to the effective potential
\begin{gather}
    S_{-1,\beta} 
    =\sum_n\frac12 \int \frac{d^d p}{(2\pi)^d} \log\left[- (\omega_n^2 +p^2)^2 G^{-1}_\sigma(\omega_n,p) + \right. \notag\\
    \left. + \left(\omega_n^2 + p^2 \right) \left(\phi^2 + \psi^2 t^2\right)  \right]
\end{gather}
Which leads to a unique thermal vacuum
with $\phi^2 = 0$ and, for negative $t$,
\begin{gather}
    - t \psi^2 = \frac{\Gamma\left(\frac{d-2}{2}\right)\zeta(d-2)}{4 \pi^\frac{d}{2} \beta^{d-2}}
    %\frac{\pi^\frac{d-5}{2}}{2 \beta^{d-2}} \zeta (3-d) \Gamma
   %\left(\frac{3-d}{2}\right)~.
\end{gather}
In summary, for $3<d<4$ the fixed point at large finite $N$ with $t = -\frac{(d-1)(d-2)}{2}$ is unitary, has a unique vacuum at zero temperature, and at finite temperature has a nonzero VEV for $\psi^2$ and hence breaks $\mathbb{Z}_2$ symmetry spontaneously at any temperature.  Note that the equation \eqref{eq:largeNeq34d} admits a solution with 
$\sigma \neq 0$ and 
$\psi = \phi = 0$. However, one can verify that this solution corresponds to a higher free energy, implying that the system favors a spontaneously broken phase with 
$\phi \neq 0$. Let us note that while such a model can be realized on a lattice, the corresponding target space is non-compact, which in turn obstructs the restoration of symmetries at high temperatures \cite{Han:2025eiw}.

Let us now finally consider the case of most interest, $d=3$, 
where we also need to take care of the $\psi^6$ interaction as in~\eqref{gcouplingi}. 
The action at the fixed point with $t=-1$ is 
\begin{gather}
    S = \int d^3 x \left[\frac12 \left(\partial_\mu \phi_i\right)^2 + \frac12 \sigma \phi_i^2  + \right. \notag \\ \left.
    \frac12 \left(\partial_\mu \psi\right)^2 -  \frac12 \sigma \psi^2 + \frac{g_*}{6! N^2} \psi^6 \right]
\end{gather}
and $g_* = 15360$.
Again integrating out $N-1$ degrees of freedoms we arrive at 
\begin{gather}
    S = \frac{N-1}{2} \tr \log \left(-\Delta + \sigma\right) + \\
    \int d^3 x \left[\frac12 \left(\partial_\mu \phi\right)^2 + \frac12 \sigma \phi^2  + \frac12 \left(\partial_\mu \psi\right)^2 - \frac12 \sigma \psi^2 + \frac{g_*}{6! N^2} \psi^6 \right] \notag
\end{gather}
Minimizing with respect to $\phi$ and $\psi$, and rescaling $\phi,\psi\to N^{1/2} \phi,N^{1/2}\psi$ we arrive at
\begin{gather}
   \sigma \phi = 0, \quad \sigma \psi = \frac{g_*}{5!} \psi^5, \quad
   \frac{1}{8\pi}\sigma^\frac{1}{2} - \frac12 \phi^2 + \frac12  \psi^2 = 0
\end{gather}
Here the situation is different from the previous case. Because of the presence of $\psi^6$ we immediately resolve the zero-temperature manifold of vacua. So at zero temperature we get $\psi = \phi = 0$.
Let us discuss the finite temperature case. We get the following saddle point equations:
\begin{gather}
    \sigma \phi = 0, \quad -\sigma \psi + \frac{g_*}{5!} \psi^5 = 0, \notag\\
  \left(\frac{1}{-\Delta + \sigma}\right)_{x,x}+  \phi^2 - \psi^2 = 0
\end{gather}
We find that the thermal vacuum is at $\phi = 0$ and $\psi \neq 0 $, determined by the following equations
\begin{gather}
   \psi^2 =  \left(\frac{1}{-\Delta + \sigma}\right)_{x,x} = -\frac{\log \left(2 \sinh \left(\frac{\beta  \sqrt{\sigma }}{2}\right)\right)}{2 \pi \beta}, \, \sigma = \frac{g_*}{5!} \psi^4. \notag
   % \\
   % \psi^2 \beta = - \frac{1}{2\pi} \log  \left(2 \sinh \left( \sqrt{\frac{g_*}{4 \cdot 5!}}\psi^2 \beta \right)\right), \notag
\end{gather}
Plugging $g_* = 15360$ we find that the thermal vacuum is approximately at  $\psi^2 = \frac{0.05963}{\beta}$.
Therefore, in this $2+1$ dimensional theory, which appears to be healthy on all accounts, there is symmetry breaking at arbitrary $\beta$. Since this fixed point is multi-critical, we can infer that in some slices of the phase diagram there is order at infinite temperature, but not necessarily everywhere. Since the multi-critical point admits gapped deformations, we can also conclude that some gapped disordered phases become ordered as we crank up the temperature and stay ordered forever.

\section*{Acknowledgements}

We thank Albert Schwartz, Andrew Lucas and Soumyadeep Chaudhuri for insightful discussion and correspondence. We also thank Misha Smolkin, Yifan Wang and Igor Klebanov for their valuable comments on the draft.

\bibliographystyle{unsrt}
\bibliography{bibliography}

\newpage
\onecolumngrid
\appendix

\section{Computation of $\beta$ function for $\psi^6$ coupling}\label{A}

Let us compute the beta function for the coupling $\psi^6$ in 3 dimensions. The propagator of $\psi$ and $\sigma$ field in momentum space is \cite{Moshe:2003xn}
\begin{gather}
    G_\psi = \frac{1}{p^2}, \quad G_\sigma = - 16 p,
\end{gather}
% We are going to use the following formula
% \begin{gather}
%     \int \frac{d^d p}{(2\pi)^d} \frac{e^{i p x}}{(p^2)^\alpha} = \frac{1}{2^{2\alpha} \pi^\frac{d}{2}} \frac{\Gamma\left(\frac{d}{2} - \alpha\right)}{\Gamma\left(\alpha\right)}  \frac{1}{(x^2)^{\left(\frac{d}{2}-\alpha\right)}}
% \end{gather}
% For that we need the following formula. The propagator for $\psi$ field is
% \begin{gather}
%     G_{\psi} = \frac{1}{4\pi |x|} = \int \frac{d^3 p}{(2\pi)^3} \frac{e^{i p x}}{p^2}, \notag\\
%     G^2_\psi  = \frac{1}{16 \pi^2 x^2}, \quad \left(G^2_\psi\right)(p) = \frac{1}{8 p}, \notag\\
%     G_\sigma =- 2(G^2)^{-1} = - 16 p
% \end{gather}
% With that knowledge we can start computing the diagrams.
Schematically, to the leading order in $\frac{1}{N}$ expansion we get the following contributions to the sextic correlator of $\psi$ fields
\begin{gather} 
\langle \psi(x_1) \psi(x_2) \psi(x_3) \psi(x_4) \psi (x_5) \psi (x_6) \rangle = 
\frac{1}{6!} \vcenter{\hbox{
    \begin{tikzpicture}[scale = 0.5]
        \draw (-2,0) -- (2,0);
        \draw (-1,1.72) -- (1,-1.72);
        \draw (-1,-1.72) -- (1,1.72);
    \end{tikzpicture}
    }} + \frac{1}{48}\vcenter{
\hbox{
\begin{tikzpicture}
\draw (-0.75,-0.25) -- (-0.5,0) -- (-0.75,0.25);
\draw (0.75,-0.25) -- (0.5,0) -- (0.75,0.25);
\draw[dashed] (-0.5,0) -- (0.5,0); 
\draw[dashed] (0,0) -- (0,-0.5);
\draw (-0.5,-1) -- (0,-0.5) -- (0.5,-1);
\draw[fill=black!20] (0,0) circle (0.25);
\end{tikzpicture}
}} +  \notag\\
    \frac{15}{6!} \vcenter{\hbox{
    \begin{tikzpicture}[scale = 0.5]
        \draw (-2,0) -- (2,0);
        \draw (-1,1.72) -- (1,-1.72);
        \draw[dashed] (-0.5,0.86) -- (0.5,0.86);
        \draw (-1,-1.72) -- (1,1.72);
    \end{tikzpicture}
    }} + \frac{6}{6!} \vcenter{\hbox{
    \begin{tikzpicture}[scale = 0.5]
        \draw (-2,0) -- (2,0);
        \draw (-1,1.72) -- (1,-1.72);
        \draw (-1,-1.72) -- (1,1.72);
        \draw[dashed] (0.5,0) .. controls (1,1) .. (1.5,0);
    \end{tikzpicture}
    }} + \frac16 \vcenter{\hbox{
    \begin{tikzpicture}[scale = 0.5]
    \draw (-2,0) -- (-1,0);
    \draw[dashed] (-0.5,0.86) -- (0.5,0.86);
    \draw (0.5,0.86) -- (1,0);
    \draw[dashed] (0.5,-0.86) -- (1,0);
    \draw (-0.5,-0.86) -- (0.5,-0.86);
    \draw[dashed] (-0.5,-0.86) -- (-1,0);
    \draw (-0.5,0.86) -- (-1,0);
    \draw (2,0) -- (1,0);
    \draw (-1,1.72) -- (-0.5,0.86);
    \draw (1,-1.72) -- (0.5,-0.86);
    \draw (-1,-1.72) -- (-0.5,-0.86);
    \draw (0.5, 0.86)-- (1,1.72);
\end{tikzpicture}
    }}
    + \frac{1}{4}  \vcenter{\hbox{
\begin{tikzpicture}
\draw (-0.75,-0.25) -- (-0.5,0) -- (-0.75,0.25);
\draw[dashed] (0,0) -- (0.5,0.5); % Upper-left dashed line
\draw[dashed] (0,0) -- (-0.5,0); % Bottom dashed line
\draw[dashed] (0,0) -- (0.5,-0.5); % Upper-right dashed line
\draw[fill=black!20] (0,0) circle (0.25); % Circle at the center
\draw (0.25,0.75) -- (0.5,0.5) -- (1.0,0.5) -- (1.25,0.75);
\draw[dashed] (1.0,0.5)--(1.0,-0.5);
\draw (0.25,-0.75) -- (0.5,-0.5) -- (1.0,-0.5) -- (1.25,-0.75);
\end{tikzpicture}}}
+     \frac{1}{8}\vcenter{\hbox{
\begin{tikzpicture} % Upper-left dashed line
\draw (-0.75,-0.25) -- (-0.5,0) -- (-0.75,0.25);
\draw (1.75,-0.25) -- (1.5,0) -- (1.75,0.25);
\draw[dashed] (-0.5,0) -- (1.5,0); % Bottom dashed line
\draw[dashed] (0,0) -- (0,-0.5); 
\draw[dashed] (1,0) -- (1,-0.5);% % Upper-right dashed line
\draw[fill=black!20] (0,0) circle (0.25);
\draw[fill=black!20] (1,0) circle (0.25);
\draw (-0.25,-0.75) -- (0,-0.5) -- (1.0,-0.5) -- (1.25,-0.75);
\end{tikzpicture}}} 
+ \label{app:eq:log1}\\ + \frac{1}{48}\vcenter{\hbox{
\begin{tikzpicture}
\draw (-0.75,-0.25) -- (-0.5,0) -- (-0.75,0.25);
\draw (1.75,-0.25) -- (1.5,0) -- (1.75,0.25);
\draw[dashed] (-0.5,0) -- (1.5,0); % Upper-left dashed line
\draw[dashed] (0,0) -- (-0.5,0); % Bottom dashed line
\draw[dashed] (0,0) -- (0.5,-0.5);
\draw[dashed] (0.5,-0.5) -- (1,0); % Upper-right dashed line
\draw[dashed] (0.5,-0.5) -- (0.5,-1);
\draw (0.25,-1.25) -- (0.5,-1) -- (0.75,-1.25);
\draw[fill=black!20] (0,0) circle (0.25);
\draw[fill=black!20] (1,0) circle (0.25);
\draw[fill=black!20] (0.5,-0.5) circle (0.25);% Circle at the center
\end{tikzpicture}}}  
+ \frac{1}{32}
\vcenter{
\hbox{
\begin{tikzpicture}
\draw (-0.75,-0.25) -- (-0.5,0) -- (-0.75,0.25);
\draw (1.75,-0.25) -- (1.5,0) -- (1.75,0.25);
\draw[dashed] (-0.5,0) -- (1.5,0); 
\draw[dashed] (0.25,0) circle (0.25);
\draw[dashed] (1,0) -- (1,-0.5);
\draw (0.75,-0.75) -- (1,-0.5) -- (1.25,-0.75);
\draw[fill=black!20] (1,0) circle (0.25);
\draw[fill=black!20] (0,0) circle (0.25);
\end{tikzpicture}
}}
+ \frac{1}{32}
\vcenter{
\hbox{
\begin{tikzpicture}
\draw (-0.75,-0.25) -- (-0.5,0) -- (-0.75,0.25);
\draw (1.75,-0.25) -- (1.5,0) -- (1.75,0.25);
\draw[dashed] (-0.5,0) -- (1.5,0); 
\draw[dashed] (0,0.25) circle (0.25);
\draw[dashed] (1,0) -- (1,-0.5);
\draw (0.75,-0.75) -- (1,-0.5) -- (1.25,-0.75);
\draw[fill=black!20] (1,0) circle (0.25);
\draw[fill=black!20] (0,0) circle (0.25);
\end{tikzpicture}
}}
+ 
\frac{1}{16}\vcenter{
\hbox{
\begin{tikzpicture}
 \draw (-0.75,-0.25) -- (-0.5,0) -- (-0.75,0.25);
\draw (0.75,-0.25) -- (0.5,0) -- (0.75,0.25);
\draw[dashed] (-0.5,0) -- (0.5,0); 
\draw[dashed] (0,0.25) circle (0.25);
\draw[dashed] (0,0) -- (0,-0.5);
\draw (-0.25,-0.75) -- (0,-0.5) -- (0.25,-0.75);
\draw[fill=black!20] (0,0) circle (0.25);
\end{tikzpicture}
}} +  \frac{1}{16}
\vcenter{
\hbox{
\begin{tikzpicture}
\draw (-0.75,-0.25) -- (-0.5,0) -- (-0.75,0.25);
\draw (1.0,-0.25) -- (0.75,0) -- (1.0,0.25);
\draw[dashed] (-0.5,0) -- (0.75,0); 
\draw[dashed] (0.25,0) circle (0.2);
\draw[dashed] (0,0) -- (0,-0.5);
\draw (-0.25,-0.75) -- (0,-0.5) -- (0.25,-0.75);
\draw[fill=black!20] (0,0) circle (0.25);
\end{tikzpicture}
}} 
+ \label{app:eq:log2}\\ + \frac{1}{16} \vcenter{
\hbox{
\begin{tikzpicture}
\draw (-0.75,-0.25) -- (-0.5,0) -- (-0.75,0.25);
\draw (0.75,-0.25) -- (0.5,0) -- (0.75,0.25);
\draw[dashed] (-0.5,0) -- (0.5,0); 
\draw[dashed] (0,0) -- (0,-0.5);
\draw (-0.5,-1) -- (0,-0.5) -- (0.5,-1);
\draw[dashed] (-0.25,-0.75) -- (0.25,-0.75);
\draw[fill=black!20] (0,0) circle (0.25);
\end{tikzpicture}
}}
+  \frac{1}{8} \vcenter{
\hbox{
\begin{tikzpicture}
\draw (-0.75,-0.25) -- (-0.5,0) -- (-0.75,0.25);
\draw (0.75,-0.25) -- (0.5,0) -- (0.75,0.25);
\draw[dashed] (-0.5,0) -- (0.5,0); 
\draw[dashed] (0,0) -- (0,-0.5);
\draw (-0.5,-1) -- (0,-0.5) -- (0.5,-1);
\draw[dashed] (0.125,-0.625) .. controls (0.375,-0.5).. (0.375,-0.875);
\draw[fill=black!20] (0,0) circle (0.25);
\end{tikzpicture}
}} \label{app:eq:diagrams} + \\
+   \vcenter{\hbox{
    \begin{tikzpicture}
    \draw (-1.25,-0.25) -- (-1,0) -- (-1.25,0.25);
    \draw (2.25,-0.25) -- (2,0) -- (2.25,0.25);
    \draw[dashed] (-1,0) -- (-0.25,0);
    \draw[dashed] (0.75,0) -- (2,0);
    \draw[dashed] (0.25,0) circle (0.5);
    \draw[dashed] (1.5,0) -- (1.5,-0.5);
    \draw (1.25,-0.75) -- (1.5,-0.5) -- (1.75,-0.75);
    \draw[fill=black!20] (1.5,0) circle (0.25);
    \draw[fill=black!20] (0.75,0) circle (0.125);
    \draw[fill=black!20] (-0.25,0) circle (0.125);
    \end{tikzpicture}
    }} +
% \end{gather}
% We should also be concerned about these two diagrams
% \begin{gather}
    \frac{1}{8}\vcenter{\hbox{
\begin{tikzpicture} % Upper-left dashed line
\draw (-0.75,-0.25) -- (-0.5,0) -- (-0.75,0.25);
\draw (1.75,-0.25) -- (1.5,0) -- (1.75,0.25);
\draw[dashed] (-0.5,0) -- (1.5,0); % Bottom dashed line
\draw[dashed] (0.5,0) -- (0,-0.5); 
\draw[dashed] (0.5,0) -- (1,-0.5);% % Upper-right dashed line
\draw[fill=black!20] (0.5,0) circle (0.25);
\draw (-0.25,-0.75) -- (0,-0.5) -- (1.0,-0.5) -- (1.25,-0.75);
\end{tikzpicture}}} 
+ \frac{1}{48}\vcenter{\hbox{
\begin{tikzpicture}
\draw (-0.75,-0.25) -- (-0.5,0) -- (-0.75,0.25);
\draw (1.75,-0.25) -- (1.5,0) -- (1.75,0.25);
\draw[dashed] (0.5,0) -- (1.5,0); % Upper-left dashed line
\draw[dashed] (0.5,0) -- (-0.5,0); % 
% \draw[dashed] (0,0) -- (-0.5,0); % Bottom dashed line
% \draw[dashed] (0,0) -- (0.5,-0.5);
% \draw[dashed] (0.5,-0.5) -- (0.5,-1);
% \draw (0.25,-1.25) -- (0.5,-1) -- (0.75,-1.25);
\draw[dashed] (0.5,0) .. controls (0.75,-0.5) .. (0.5,-1);
\draw[dashed] (0.5,0) .. controls (0.25,-0.5) .. (0.5,-1);
\draw[dashed] (0.5,-1.5)--(0.5,-1);
\draw[fill=black!20] (0.5,0) circle (0.25);
\draw[fill=black!20] (0.5,-1) circle (0.25);% Circle at the center
\draw (0.75,-1.75) -- (0.5,-1.5) -- (0.25,-1.75);
\end{tikzpicture}}}  \label{app:eq:lineardiv}
\end{gather}
The logarithmic divergences arise from the lines \eqref{app:eq:log1}--\eqref{app:eq:diagrams}, and the linear divergences appear in \eqref{app:eq:lineardiv}. One can check that the linear divergences are canceled.
Now let us consider diagrams from \eqref{app:eq:diagrams}
\begin{gather*}
    \vcenter{
    \hbox{
    \begin{tikzpicture}[scale = 0.5]
        \draw (-2,0) -- (2,0);
        \draw (-1,1.72) -- (1,-1.72);
        \draw[dashed] (-0.5,0.86) -- (0.5,0.86);
        \draw (-1,-1.72) -- (1,1.72);
    \end{tikzpicture}
    }} = \frac{(-g)(-t)^2}{N^3} \int \frac{d^3 q}{(2\pi)^3} \frac{( - 16 q)} {q^4} =  \frac{8 g t^2}{\pi^2 N^3} \log \Lambda, \, 
\vcenter{\hbox{
    \begin{tikzpicture}[scale = 0.5]
    \draw (-2,0) -- (-1,0);
    \draw[dashed] (-0.5,0.86) -- (0.5,0.86);
    \draw (0.5,0.86) -- (1,0);
    \draw[dashed] (0.5,-0.86) -- (1,0);
    \draw (-0.5,-0.86) -- (0.5,-0.86);
    \draw[dashed] (-0.5,-0.86) -- (-1,0);
    \draw (-0.5,0.86) -- (-1,0);
    \draw (2,0) -- (1,0);
    \draw (-1,1.72) -- (-0.5,0.86);
    \draw (1,-1.72) -- (0.5,-0.86);
    \draw (-1,-1.72) -- (-0.5,-0.86);
    \draw (0.5, 0.86)-- (1,1.72);
\end{tikzpicture}
    }} = \frac{(-t)^6}{N^3} \int \frac{d^3 q}{(2\pi)^3} \frac{(-16)^3}{q^3} = -\frac{2048 t^6}{\pi^2 N^3} \log \Lambda, \notag\\
 \vcenter{\hbox{
\begin{tikzpicture}
\draw (-0.75,-0.25) -- (-0.5,0) -- (-0.75,0.25);
\draw[dashed] (0,0) -- (0.5,0.5); % Upper-left dashed line
\draw[dashed] (0,0) -- (-0.5,0); % Bottom dashed line
\draw[dashed] (0,0) -- (0.5,-0.5); % Upper-right dashed line
\draw[fill=black!20] (0,0) circle (0.25); % Circle at the center
\draw (0.25,0.75) -- (0.5,0.5) -- (1.0,0.5) -- (1.25,0.75);
\draw[dashed] (1.0,0.5)--(1.0,-0.5);
\draw (0.25,-0.75) -- (0.5,-0.5) -- (1.0,-0.5) -- (1.25,-0.75);
\end{tikzpicture}}} = \frac{(-2)(-1)^3(-t)^5}{N^3} \int \frac{d^3 q}{(2\pi)^3} \frac{(-16)^3}{q^3}, \quad
\vcenter{\hbox{
\begin{tikzpicture} % Upper-left dashed line
\draw (-0.75,-0.25) -- (-0.5,0) -- (-0.75,0.25);
\draw (1.75,-0.25) -- (1.5,0) -- (1.75,0.25);
\draw[dashed] (-0.5,0) -- (1.5,0); % Bottom dashed line
\draw[dashed] (0,0) -- (0,-0.5); 
\draw[dashed] (1,0) -- (1,-0.5);% % Upper-right dashed line
\draw[fill=black!20] (0,0) circle (0.25);
\draw[fill=black!20] (1,0) circle (0.25);
\draw (-0.25,-0.75) -- (0,-0.5) -- (1.0,-0.5) -- (1.25,-0.75);
\end{tikzpicture}}} = \frac{(-2)^2 (-1)^6 (-t)^4}{N^3} \int \frac{d^3 q}{(2\pi)^3} \frac{(-16)^3}{q^3}, \notag\\
\vcenter{\hbox{
\begin{tikzpicture}
\draw (-0.75,-0.25) -- (-0.5,0) -- (-0.75,0.25);
\draw (1.75,-0.25) -- (1.5,0) -- (1.75,0.25);
\draw[dashed] (-0.5,0) -- (1.5,0); % Upper-left dashed line
\draw[dashed] (0,0) -- (-0.5,0); % Bottom dashed line
\draw[dashed] (0,0) -- (0.5,-0.5);
\draw[dashed] (0.5,-0.5) -- (1,0); % Upper-right dashed line
\draw[dashed] (0.5,-0.5) -- (0.5,-1);
\draw (0.25,-1.25) -- (0.5,-1) -- (0.75,-1.25);
\draw[fill=black!20] (0,0) circle (0.25);
\draw[fill=black!20] (1,0) circle (0.25);
\draw[fill=black!20] (0.5,-0.5) circle (0.25);% Circle at the center
\end{tikzpicture}}} = \frac{(-2)^3 (-1)^9 (-t)^3}{N^3} \int \frac{d^3 q}{(2\pi)^3} \frac{(-16)^3}{q^3}
\end{gather*}
The other diagrams could be computed in the same fashion. At the end we get the following beta-function
\begin{gather}
    \beta_g = \frac{128 t^2 g}{N \pi^2} - \frac{245760}{N \pi^2} t^3(t-1)^3 
\end{gather}

\end{document}